\documentstyle[12pt,rotating]{article}
\begin{document}
\renewcommand{\thefootnote}{\fnsymbol{footnote}}
\thispagestyle{empty}
\vspace*{-2 cm}
\hspace*{\fill}  \mbox{WUE-ITP-96-009}\\
\hspace*{\fill}  \mbox{hep-ph/9606291}\\
\vspace*{2 cm}
\begin{center}
{\Large \bf Neutralino mass bounds at the
upgraded LEP collider}
\\ [3 ex] 
{\large F.~Franke\footnote{email: fabian@physik.uni-wuerzburg.de},
S. Hesselbach\footnote{email: hesselb@physik.uni-wuerzburg.de}\\ [2 ex]
Institut f\"ur Theoretische Physik, Universit\"at W\"urzburg \\
D-97074 W\"urzburg, Germany}
\end{center}
\vfill

{\bf Abstract}

Assuming that no supersymmetric signature will be found at
the upgraded LEP collider 
we derive lower bounds on the masses of the four neutralinos 
$\tilde{\chi}_i^0$ in
the Minimal Supersymmetric Standard Model (MSSM).
We consider the recently published results from the search for the
light chargino $\tilde{\chi}^\pm_1$ and the next-to-lightest
neutralino $\tilde{\chi}^0_2$ at LEP1.5 and study the
consequences of possible future
lower $\tilde{\chi}^\pm_1$ and $\tilde{\chi}^0_2$
mass limits between 65 and 95 GeV.
For a chargino mass bound of 66.8 GeV at LEP1.5, a massless
neutralino is not excluded for $\tan\beta<1.2$. If either
$\tan\beta > 2.3$ or the gluino mass $m_{\tilde{g}}>160$ GeV, we find
$m_{\tilde{\chi}_1^0} > 28$ GeV.
A possible chargino bound $m_{\tilde{\chi}_1^\pm} > 95$ GeV at LEP2
would raise this bound to 31 GeV (for all $\tan\beta$) or 
44 GeV ($\tan\beta > 2$).

\vfill
\begin{center}
May 1996
\end{center}

\newpage

\section{Introduction}
The upgrading of the LEP collider which has already started with an increase
of the available center-of-mass energy from $\sqrt{s} \approx m_Z$ 
(LEP1) to
136 GeV (LEP1.5) and will continue up to $\sqrt{s}\approx 190$ GeV  
(LEP2) in 1997
opens fascinating opportunities for precision tests of the
standard model (SM) as well as for detecting first signatures of new physics. 
It is widely acknowledged that 
supersymmetry (SUSY) \cite{susy} is the most likely theory beyond the SM.
Therefore the search for 
supersymmetric particles plays a fundamental role at the present and
future high-energy colliders and also in the program of the
upgraded LEP \cite{lep2rep}. Until now, however, no
direct evidence for SUSY has been found. Therefore the experiments at LEP and
TEVATRON resulted in lower mass limits
for SUSY particles. 
At the upgraded LEP, one expects either 
the spectacular identification of a SUSY particle
or the lower mass bounds will increase. Under the assumption 
that the production of SUSY particles is kinematically allowed,  
there exist several channels
for the detection of a supersymmetric signature.
Among the most promising processes is the 
pair production of neutralinos 
or charginos, the mass eigenstates of the fermionic partners of
the gauge and Higgs bosons.

In the present paper we use the {\it Minimal Supersymmetric Standard 
Model} (MSSM) \cite{mssm} as framework for the calculation of
the neutralino mass reach to be probably covered at the upgraded LEP
collider. In the MSSM one of the charged superpartners is expected to
be the first SUSY particle to be detected or excluded  up to a mass of
$\sim\sqrt{s}/2$. Mainly the light chargino 
$\tilde{\chi}_1^\pm$ and a light scalar top quark 
$\tilde{t}_1$ are discussed to be the lightest visible supersymmetric
particle \cite{lep2rep}.
In order to estimate the chances
for a supersymmetric signature one has to analyze carefully the
possible decay channels. Due to R-parity conservation in the
MSSM, all decay products contain the invisible lightest 
supersymmetric particle (LSP) which is assumed to be the
lightest neutralino $\tilde{\chi}_1^0$. 
In the case of stop and chargino production,
there are also charged quarks or leptons in the final state 
which could lead to a clear signature identifying or ruling out
the respective SUSY particle. The production cross sections
for stops \cite{bartlstops} and charginos \cite{bartlchar}
and their branching ratios 
as a function of the supersymmetric parameters are well-known
for LEP2 energies and form the theoretical basis for
the experimental search at LEP.

Therefore the first results of the LEP1.5 run with $\sqrt{s}=136$ GeV  
\cite{lep15, delphi15, aleph15, l315, opal15} all contain 
a discussion of the chargino search. Since no supersymmetric
signature was found, the LEP collaborations have reported a lower 
limit on the mass of the light chargino $m_{\tilde{\chi}_1^\pm} > 65$~GeV
with some dependence on the chargino mixing, the
sneutrino mass and the mass difference to the
LSP. The DELPHI collaboration \cite{delphi15} has set a preliminary limit of
$m_{\tilde{\chi}_1^\pm} > 66.8 \: \mbox{GeV}$
for $m_{\tilde{\chi}_1^\pm} - m_{\tilde{\chi}_1^0} > 10$ GeV
and 
$m_{\tilde{\chi}_1^\pm} > 63.8 \: \mbox{GeV}$
for $m_{\tilde{\chi}_1^\pm} - m_{\tilde{\chi}_1^0} = 5$~GeV
and a sneutrino mass of 1~TeV,
while ALEPH \cite{aleph15} found a lower mass bound of 67.8 GeV for
gaugino-like charginos and the sneutrino heavier than 200~GeV, or
65~GeV for a higgsino-like chargino when the mass difference to the LSP
is larger than 10~GeV. Finally OPAL \cite{opal15} 
derived lower chargino mass bounds
between 60.7 (58.7)~GeV for the smallest possible universal scalar mass $m_0$
and 65.4 (65.6)~GeV for $m_0>1$~TeV and $\tan\beta=1.5\; (35)$,
again with the mass difference condition 
$m_{\tilde{\chi}_1^\pm} - m_{\tilde{\chi}_1^0} > 10$~GeV. 

Another candidate for the lightest visible supersymmetric particle is
the second lightest neutralino 
$\tilde{\chi}_2^0$ which can be identified by its decay into the LSP.
The dominant decay channels, however, significantly depend on the
neutralino mixing and vary within different regions of the parameter
space \cite{ambrosanio}. 
With the results of the LEP1.5 run,
the ALEPH and OPAL collaborations have set limits
on the cross sections $\sigma (e^+e^- \rightarrow 
\tilde{\chi}_1^0 \tilde{\chi}_2^0)$ as a function of the neutralino masses
\cite{aleph15, opal15}. For a higgsino-like $\tilde{\chi}_2^0$ 
and $m_{\tilde{\chi}_2^0}-m_{\tilde{\chi}_1^0} > 10$~GeV
ALEPH found a lower
mass bound for the second lightest neutralino of
$m_{\tilde{\chi}_2^0} > 69 \: \mbox{GeV}$.
The most detailed neutralino mass bounds at LEP1.5 are derived by the
OPAL collaboration. Their $\tilde{\chi}_1^0$ ($\tilde{\chi}_2^0$) bounds
range from 12.0 (45.3)~GeV for a minimal universal scalar mass $m_0$ and
$\tan\beta=1.5$ to 35.2 (67.5)~GeV for $m_0=1$~TeV and $\tan\beta=35$
with a mass difference between the two light neutralinos larger than 10~GeV.
But also OPAL does not yet study precisely the dependence on $\tan\beta$.

In this paper, however, we consider general neutralino mixing,
scan over a wide theoretically acceptable parameter range and do not
impose any restrictions on the mass difference to the LSP or mixing types.
We mainly pursue two aims: First we want to analyze the lower mass bounds 
for all
four neutralinos derived from the LEP1.5 results.
Here, we also consider the effect of the TEVATRON constraints 
on the gluino mass
bounding the SUSY parameter $M$.
Second, we want to study the development of
the bounds on the way to LEP2 if no neutralino or chargino is found.
Therefore we consider possible future chargino mass bounds up to 95~GeV
and also the consequences of such a bound for the
second lightest neutralino. 
Similar studies for LEP1 were performed e.g. in \cite{roszkowski}.

In our analysis we have to take into account that 
the masses and mixings of charginos \cite{bartlchar}
and neutralinos \cite{bartlneu} are strongly correlated. Both mixing 
matrices depend on the same parameters, namely the $SU(2)$ and $U(1)$ 
gaugino masses
$M'$ and $M$, which are connected by the usual
GUT relation $M'/M=5/3\tan^2\theta_W$,
the $\mu$ parameter in the superpotential and the ratio
of the vacuum expectation values of the Higgs doublets 
$\tan\beta=v_2/v_1$. Therefore new chargino mass bounds also
result in lower mass bounds for the neutralinos even without
considering experimental constraints in the neutralino
sector. In fact it turns out that the constraints from
negative chargino search represent the by far stronger
criterium for the exclusion of a parameter region.
Only a small domain is additionally excluded by 
neutralino constraints.
In view of the further increase of the LEP energy towards LEP2
we therefore study the consequences for
the neutralino mass bounds as a function of the chargino bounds.
In a second step we also include in our analysis possible
new lower limits on the mass of the second lightest
neutralino up to 95~GeV which may arise by the
neutralino search at LEP2. 
Our results make it possible to determine neutralino mass bounds
immediately when new chargino bounds are announced from a 
LEP run with increased energy.

The paper begins with a short analysis of the parameter domain
excluded by LEP1.5  
in combination with the neutralino and chargino mass contour lines
in the $(M, \mu)$ plane needed for the interpretation of the following
figures. Then we present lower mass limits for the four neutralinos
as a function of $\tan\beta$ and of prospective new mass bounds
for the light chargino and the second lightest neutralino.
Finally we explicitly give the neutralino mass limits for the cases
of $m_{\tilde{\chi}_1^\pm} > 66.8$~GeV (LEP1.5) and
$m_{\tilde{\chi}_1^\pm} > 95$~GeV (LEP2).
All results are compared with the corresponding neutralino bounds
from LEP1 \cite{lep1}.

\section{Parameter constraints}
In this paper, we start with the conservative  
LEP1.5 chargino bound of the DELPHI collaboration \cite{delphi15} 
\begin{equation}
m_{\tilde{\chi}_1^\pm} \ge 66.8 \: \mbox{GeV}
\label{charbound}
\end{equation}
and discuss the case that this bound may be raised up to 95~GeV by LEP2.
Furthermore we consider the neutralino 
constraints from LEP1 \cite{lep1} and LEP1.5 \cite{aleph15}.
In particular, we use 
\begin{enumerate}
\item the limit on the total $Z$ width
\begin{equation}
\Delta \Gamma _Z \leq 23.1 \; \mbox{MeV},
\label{total}
\end{equation}
where
\begin{equation} 
\Delta \Gamma _Z = \Gamma (Z \rightarrow \tilde{\chi}^0_i
\tilde{\chi}^0_j )+ \Gamma (Z \rightarrow \tilde{\chi}^\pm_k
\tilde{\chi}^\mp_l), 
\hspace*{0.5cm} i,j =1,\ldots 4; \; k,l=1,2; 
\end{equation}
\item the limit on the invisible $Z$ width
\begin{equation}
\Delta \Gamma_{\mbox{inv}} \leq  8.4 \; \mbox{MeV},
\label{inv}
\end{equation}
where
\begin{equation}
\Delta \Gamma_{\mbox{inv}} = \Gamma (Z \rightarrow \tilde{\chi}^0_1
\tilde{\chi}^0_1); 
\end{equation}
\item the limits from direct neutralino search at the $Z$-resonance
\begin{eqnarray}
BR (Z \rightarrow \tilde{\chi}^0_1 \tilde{\chi}^0_j )
< 2 \times 10^{-5} & & \hspace*{1cm} j=2,\ldots 4, \\
BR (Z \rightarrow \tilde{\chi}^0_i \tilde{\chi}^0_j )
< 5 \times 10^{-5} & &  \hspace*{1cm} i,j=2,\ldots 4;
\label{direct}
\end{eqnarray}
\item the limit on the cross section from direct neutralino search at LEP1.5
\begin{eqnarray}
\sigma (e^+e^- \rightarrow \tilde{\chi}^0_1 \tilde{\chi}^0_2)
< 5, \: 3, \: 1.8, \: 1.4 \; \mbox{pb}
\label{sigma}
\end{eqnarray}
as a function of the masses of the lightest and next-to-lightest neutralino
as given in ref.~\cite{aleph15}. In our calculations, we apply this limit
to all neutralino pair production channels with at least one visible
neutralino. 
\end{enumerate}
In fact one has to take into account the results from both
LEP1 and LEP1.5 since the LEP1.5 bounds of 
eqs.~(\ref{charbound}) and (\ref{sigma}) do not totally cover the
LEP1 constraints of eqs.~(\ref{total}) -- (\ref{direct}).

We will discuss the excluded parameter space and the neutralino
mass bounds in detail for the lowest possible value $\tan\beta=1$,
a small value of $\tan\beta=2$ and
a larger $\tan\beta=10$, but we will also analyze the dependence
of the neutralino mass bounds on $\tan\beta$ with the LEP1.5 and 
the possible LEP2 results. Generally, the computed mass bounds do
not significantly change for a further increased $\tan\beta>10$.
With $\tan\beta = 1$ we explicitly want to study the light neutralino
window which allowed massless neutralinos for $\tan\beta < 1.7$ at LEP1.

The excluded parameter domain in the $(M,\mu)$ 
plane for $\tan\beta=1,2,10$ from the $Z$ width measurements and the
direct neutralino and chargino search at LEP1 and LEP1.5 is shown in Fig.~1.
For all our calculations of mass bounds in this paper we
consider the SUSY parameter range $0<M<400$~GeV and 
$-500$~GeV~$< \mu<500$~GeV of Fig.~1.  
The parameter regions that are excluded by the negative
chargino search at LEP1.5 and that may be covered at LEP2 are
marked by the contour lines for a chargino mass 
of 66.8~GeV and 95~GeV, respectively.
The values of $\tan\beta$ in Fig.~1 represent the three different
cases how this parameter domain is extended by direct neutralino
searches at LEP1 and LEP1.5. For $\tan\beta = 1$ direct neutralino
searches at both LEP energies lead to exclusively excluded parameter
regions additionally to the chargino search (Fig.~1a). Nevertheless
massless neutralinos remain allowed for small parameters $|\mu|$ and
$M$. This light neutralino window
depending on $\tan\beta$ will be discussed in Sec.~3. The situation is
different for increasing $\tan\beta$ where for $\tan\beta = 2$ a small
parameter region with small negative $\mu$ is excluded only by the
direct neutralino search at LEP1, while the LEP1.5 limit of
eq.~(\ref{sigma}) does not extend the parameter domain excluded by
chargino search (Fig.~1b).
For the large value $\tan\beta=10$ (Fig.~1c), the chargino bound
alone determines the excluded parameter space, since it is
stronger than the neutralino bounds from both LEP1 and LEP1.5 over the
whole $(M,\mu)$ plane. 
Therefore one may use for the calculation of the lower
neutralino mass bounds only the chargino bounds if
$\tan\beta>2$, while for smaller $\tan\beta$ also the results
from the direct neutralino search at LEP1 and LEP1.5  have to be
included.  
A LEP2 chargino bound of 95~GeV, however, would definitely improve all
parameter constraints from neutralino search at LEP1 and LEP1.5 for
all values of $\tan\beta$. 

In the following section we will discuss the neutralino masses  
which are compatible with the allowed parameter
domains. The contour lines for the mass of the lightest neutralino
shown in Fig.~1 may help to explain 
the lower mass bound for the lightest neutralino.  
For LEP1.5 and thereafter, 
this bound is mainly determined by the chargino mass limit with
the exceptions described above.

\section{Neutralino mass bounds}
In Fig.~2 the lower neutralino mass bounds are shown 
as a function of $\tan\beta$ for the new chargino bound
$m_{\tilde{\chi}^\pm_1} > 66.8$~GeV of LEP1.5 (Fig.~2a) and for a 
prospective bound $m_{\tilde{\chi}^\pm_1} > 95$~GeV after LEP2
(Fig.~2b). 
Note that we do not consider any assumptions on the mixing type or
on the mass difference between a visible neutralino and the LSP.
Therefore our bound on  
$\tilde{\chi}^0_2$ is significantly lower than the LEP1.5 bound
published by ALEPH \cite{aleph15}. 
Generally, the experimental results from LEP1.5
raise the neutralino mass limits by 5 --
10~GeV compared to the bounds from LEP1 \cite{lep1}.
But even LEP1.5 does not totally exclude a massless neutralino
for small $\tan\beta < 1.2$, while it was allowed up to
$\tan\beta <1.7$ at LEP1. One would need a lower chargino mass bound of
about 78~GeV in order to rule out a massless neutralino at LEP for all
$\tan\beta$. If no chargino will be found 
at LEP2, a lower bound $m_{\tilde{\chi}_1^0} > 31$~GeV can be expected
independently of $\tan\beta$.

A lower bound on the gluino mass 
$m_{\tilde{g}}$, however, can raise these
LEP1.5 neutralino bounds. The CDF gluino mass limits
significantly depend on the squark mass \cite{cdf}.
Assuming a heavy squark $m_{\tilde{q}}> 400$~GeV, the moderate bound 
\begin{equation}
m_{\tilde{g}}>160 \: \mbox{GeV}
\end{equation}  
restricts the parameter $M$ using the GUT relation
\begin{equation}
M=\frac{\alpha_2}{\alpha_3} \: m_{\tilde{g}} \approx 0.3 \: m_{\tilde{g}},
\end{equation}
where $\alpha_2$ and $\alpha_3$ are the gauge coupling constants of the
$SU(2)$ and $SU(3)$ gauge groups, respectively. 
We show in Fig.~2a also the lower neutralino mass bounds 
taking into account this restriction
\begin{equation}
M>50 \: \mbox{GeV}.
\end{equation}
In this case there exists a lower neutralino mass bound of 28~GeV
for all values of $\tan\beta$, a massless neutralino can be ruled out.
Generally, the LEP1.5 mass bounds for all four neutralinos are
raised compared to LEP1
for small $\tan\beta$ with this gluino mass limit, while
it has no effect for $\tan\beta>2.5$. 

In order to have also any influence on the prospective neutralino bounds at
LEP2, a gluino mass limit of at least 300~GeV is necessary
(see Fig.~1). Therefore we do not discuss such a limit in
Fig.~2b but study in Figs.~2b and 3 the impact of an experimental
lower bound for the lightest visible
neutralino $\tilde{\chi}^0_2$ of the same size as the chargino bound.
This $\tilde{\chi}^0_2$ bound serves as a rough estimate for the
possible results from direct neutralino search at LEP2 which cannot
yet be determined in detail.

Just with the chargino
limit $m_{\tilde{\chi}_1^\pm}>95$~GeV (solid lines), the lightest visible
neutralino $\tilde{\chi}_2^0$ is heavier than 54~GeV ($\tan\beta=1$) or 95~GeV
($\tan\beta=10$). Now by supposition this bound on 
$\tilde{\chi}_2^0$ is always 95~GeV in Fig.~2b (dashed lines). Then
the mass bound on $\tilde{\chi}_4^0$ is 
significantly higher for small $\tan\beta < 8$ by as much as
25~GeV. The bounds on
$\tilde{\chi}_1^0$ and $\tilde{\chi}_3^0$ are raised by about 3 -- 5~GeV
only for small $\tan\beta < 1.5$, for larger $\tan\beta$ they remain
practically unaffected by the additional $\tilde{\chi}_2^0$ bound.

Now we give an outlook on the possible increase 
of the neutralino mass limits when 
new chargino mass bounds and additional bounds on the lightest visible
neutralino arise during the further upgrading of LEP.
In Fig.~3 the neutralino bounds are depicted as a function of the
chargino limit 
for the two values $\tan\beta=2,10$. For $\tan\beta=2$ the 
$\tilde{\chi}_1^0$ ($\tilde{\chi}_2^0$) mass limit
increases from 25 (48)~GeV with $m_{\tilde{\chi}_1^\pm}
> 65$~GeV to 44 (77) GeV with $m_{\tilde{\chi}_1^\pm} > 95$~GeV. 
Larger values of $\tan\beta$ lead to a further increase of the lower
mass limits between 8~GeV for the lightest neutralino and 60~GeV
for $\tilde{\chi}_4^0$.

The dashed lines in Fig.~3 show the lower limits on the masses
of $\tilde{\chi}^0_{1,3,4}$ 
if additionally to a lower chargino mass bound also the same
mass limit for the lightest visible neutralino 
$\tilde{\chi}^0_2$ is derived by direct neutralino search.
Here for small $\tan\beta$ a similar increase of the 
$\tilde{\chi}_2^0$ and $\tilde{\chi}_4^0$ bounds occurs
as already discussed in Fig.~2, while the other neutralino mass bounds
for $\tan\beta=2$ and all bounds for $\tan\beta=10$
remain nearly unaffected. Thus for large values of $\tan\beta$
it is the chargino mass limit alone that determines the
lower neutralino mass limits at LEP2.

The neutralino bounds of Figs.~2 and 3 are summarized in Table 1 
which can easily be compared with the results of LEP1 
(for the most recent analysis of the LEP1 data see Table 3 in \cite{lep1},
earlier results can be found in \cite{lep1old}).
Except for the heaviest neutralino, there is a significant raise of the 
lower neutralino mass limits 
for $\tan\beta>1.2$. Already now after LEP1.5
the lightest neutralino 
$\tilde{\chi}_1^0$ 
must be heavier than 25~GeV compared to 20~GeV after LEP1
if $\tan\beta>2$. 
If LEP2 sets a lower chargino mass bound of 95~GeV,
the $\tilde{\chi}_1^0$
mass is larger than at least 31~GeV
(all $\tan\beta$) or 44~GeV ($\tan\beta>2$)
and a massless neutralino is excluded.
Experimental results from direct neutralino search at LEP2 may further
improve these bounds.
Including the CDF bound on the gluino mass
$m_{\tilde{g}}>160$~GeV, we obtain with the LEP1.5 results
lower mass bounds of 28 and 35~GeV for the lightest and
next-to-lightest neutralino, respectively,
independently of the value for
$\tan\beta$.

\section{Conclusion}
We have derived lower limits on the masses of the four
neutralinos if the experiments at the upgraded LEP collider
yield new chargino mass bounds between 65 and 95~GeV.
The preliminary limit $m_{\tilde{\chi}_1^\pm}>66.8$~GeV announced by
the LEP collaborations does not yet exclude massless neutralinos
for $\tan\beta<1.2$. 
A future chargino mass bound $m_{\tilde{\chi}_1^\pm}>78$~GeV, however,
would definitely rule out the existence of massless neutralinos for all
values of $\tan\beta$. 

For $\tan\beta>2$, the lower limit
on the mass of the lightest neutralino is found to be 25~GeV for the
new LEP1.5 chargino mass limit.
With the additional constraint on the gluino mass 
$m_{\tilde{g}}>160$~GeV ($M>50$~GeV)
suggested by the TEVATRON results this bound is raised to 
$m_{\tilde{\chi}_1^0} > 28$~GeV for all values of $\tan\beta$. 
If no chargino is discovered at LEP2 and a chargino mass bound
$m_{\tilde{\chi}_1^\pm}>95$~GeV is established, one gets
lower $\tilde{\chi}^0_1$ ($\tilde{\chi}^0_2$) limits of
31 (54)~GeV for all values of $\tan\beta$. 
Therefore LEP2 will
definitely find or exclude massless neutralinos in the MSSM, which are still
allowed after LEP1.5 if the gaugino mass parameter $M$ is not
constrained.

\section*{Acknowledgements}
We would like to thank H.~Fraas and A.~Bartl for the careful reading of 
the manuscript.
This work was supported by the Deutsche Forschungsgemeinschaft under
contract no.\ FR~1064/2-1. 

\newpage

\begin{table}[p]
\begin{center}
\begin{tabular}{ccrrrc}
\hline
Chargino & Neutralino & \multicolumn{3}{c}{$\tan\beta$} 
 & $m_{\tilde{g}}>160$ GeV \\ \cline{3-5}
mass bound & & $>1$ & $>2$ & $>3$ & all $\tan\beta$ \\
\hline
66.8 & $\tilde{\chi}^0_1$ & 0 & 25 & 31 & 28 \\
     & $\tilde{\chi}^0_2$ & 0 & 48 & 54 & 35 \\
     & $\tilde{\chi}^0_3$ & 75 & 84 & 89 & 83 \\ 
     & $\tilde{\chi}^0_4$ & 91 & 115 & 128 & 106 \\
\hline
95   & $\tilde{\chi}^0_1$ & 31 &  44 & 46 & 31 \\
     & $\tilde{\chi}^0_2$ & 54 &  77 & 86 & 54 \\ 
     & $\tilde{\chi}^0_3$ & 99 & 109 & 113 & 99 \\
     & $\tilde{\chi}^0_4$ & 106 & 137 & 159 & 106 \\
\hline
\end{tabular}
\end{center} 
\caption{Lower neutralino mass limits for two lower chargino mass
bounds in GeV.}
\end{table} 
\clearpage

\begin{figure}[p]
\begin{center}
\setlength{\unitlength}{1cm}
\begin{picture}(10.5,10.5)
\put(1.5,9.4){(a)}
\put(8.3,9.4){$\tan\beta=1$}
\put(-0.9,-3.1){\includegraphics{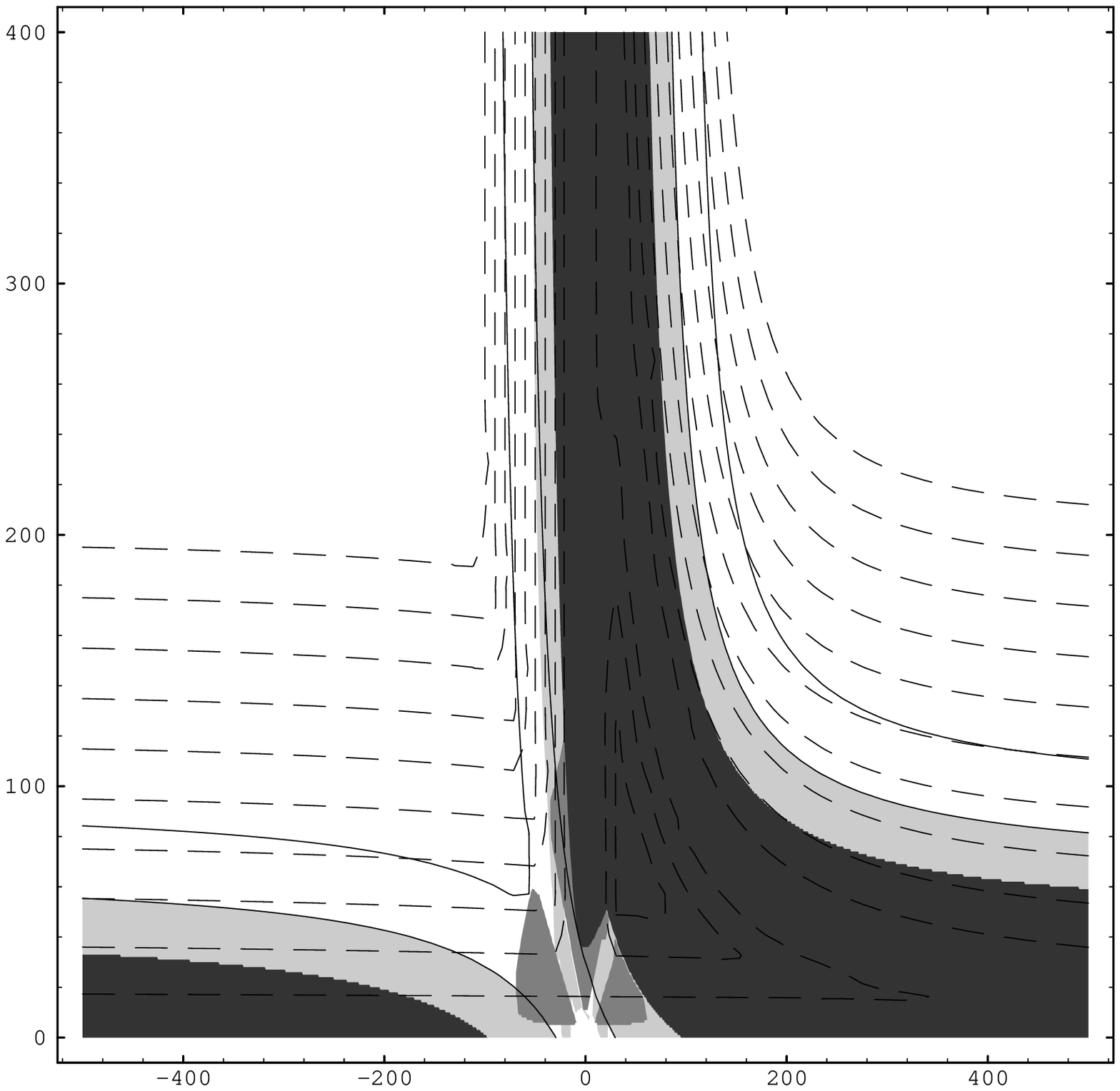}}
\put(5.2,0){$\mu/$GeV}
\put(0,4.7){\begin{sideways} $M/$GeV \end{sideways}}
\put(1.35,5.85){\scriptsize  $m_{\tilde{\chi}_1^0}/$GeV}
\put(1.35,1.77){\scriptsize 20}
\put(1.35,3.08){\scriptsize 50}
\put(1.35,5.33){\scriptsize 100}
\put(8.9,6.15){\scriptsize  $m_{\tilde{\chi}_1^\pm}/$GeV}
\put(9.7,2.78){\scriptsize 66.8}
\put(9.9,3.47){\scriptsize 95}
\end{picture}
\end{center}
Figure 1a:\ Contour lines for the mass of the lightest neutralino
(dashed; 10, 20, 30, 40, 50, 60, 70, 80, 90 and 100~GeV)
and chargino (solid; 66.8 and 95~GeV) and the excluded parameter
space from LEP1 (dark: $Z$ width
measurements, gray: direct neutralino search) and LEP1.5 (bright shaded) 
in the ($M,\mu$) plane for $\tan\beta=1$. 
\end{figure}

\begin{figure}[p]
\begin{center}
\setlength{\unitlength}{1cm}
\begin{picture}(10.5,10.5)
\put(1.5,9.4){(b)}
\put(8.3,9.4){$\tan\beta=2$}
\put(-0.9,-3.1){\includegraphics{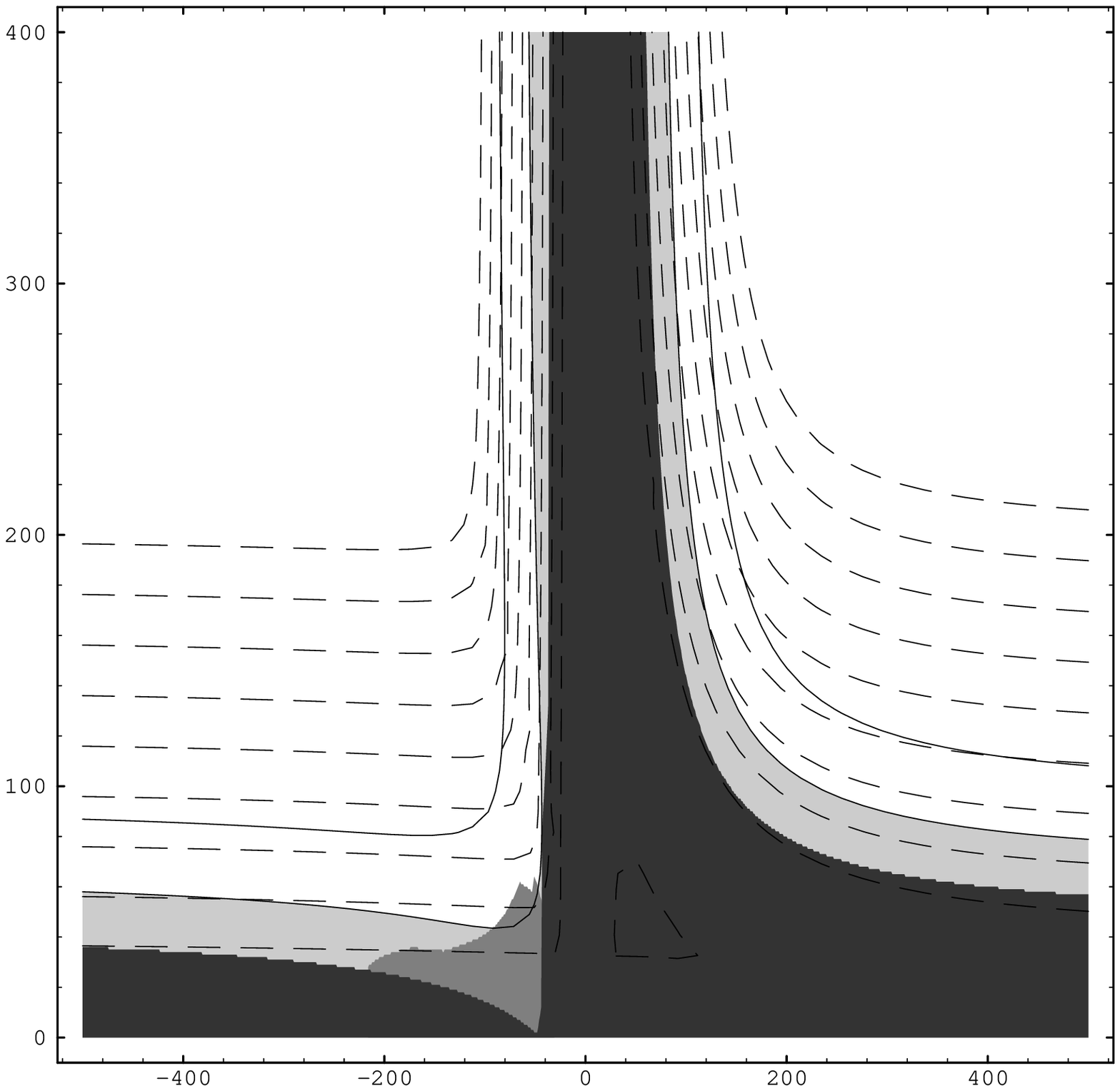}}
\put(5.2,0){$\mu/$GeV}
\put(0,4.7){\begin{sideways} $M/$GeV \end{sideways}}
\put(1.35,5.9){\scriptsize  $m_{\tilde{\chi}_1^0}/$GeV}
\put(1.35,1.77 ){\scriptsize 20}
\put(1.35,3.1){\scriptsize 50}
\put(1.35,5.35){\scriptsize 100}
\put(8.9,6.05){\scriptsize  $m_{\tilde{\chi}_1^\pm}/$GeV}
\put(9.7,2.71){\scriptsize 66.8}
\put(9.9,3.4){\scriptsize 95}
\end{picture}
\end{center}
Figure 1b:\ Contour lines for the mass of the lightest neutralino
(dashed; 20, 30, 40, 50, 60, 70, 80, 90 and 100~GeV)
and chargino (solid; 66.8 and 95~GeV) and the excluded parameter
space from LEP1 (dark: $Z$ width
measurements, gray: direct neutralino search) and LEP1.5 (bright shaded) 
in the ($M,\mu$) plane for $\tan\beta=2$. 
\end{figure}

\begin{figure}[p]
\begin{center}
\setlength{\unitlength}{1cm}
\begin{picture}(10.5,10.5)
\put(1.5,9.4){(c)}
\put(8.3,9.4){$\tan\beta=10$}
\put(-0.9,-3.1){\includegraphics{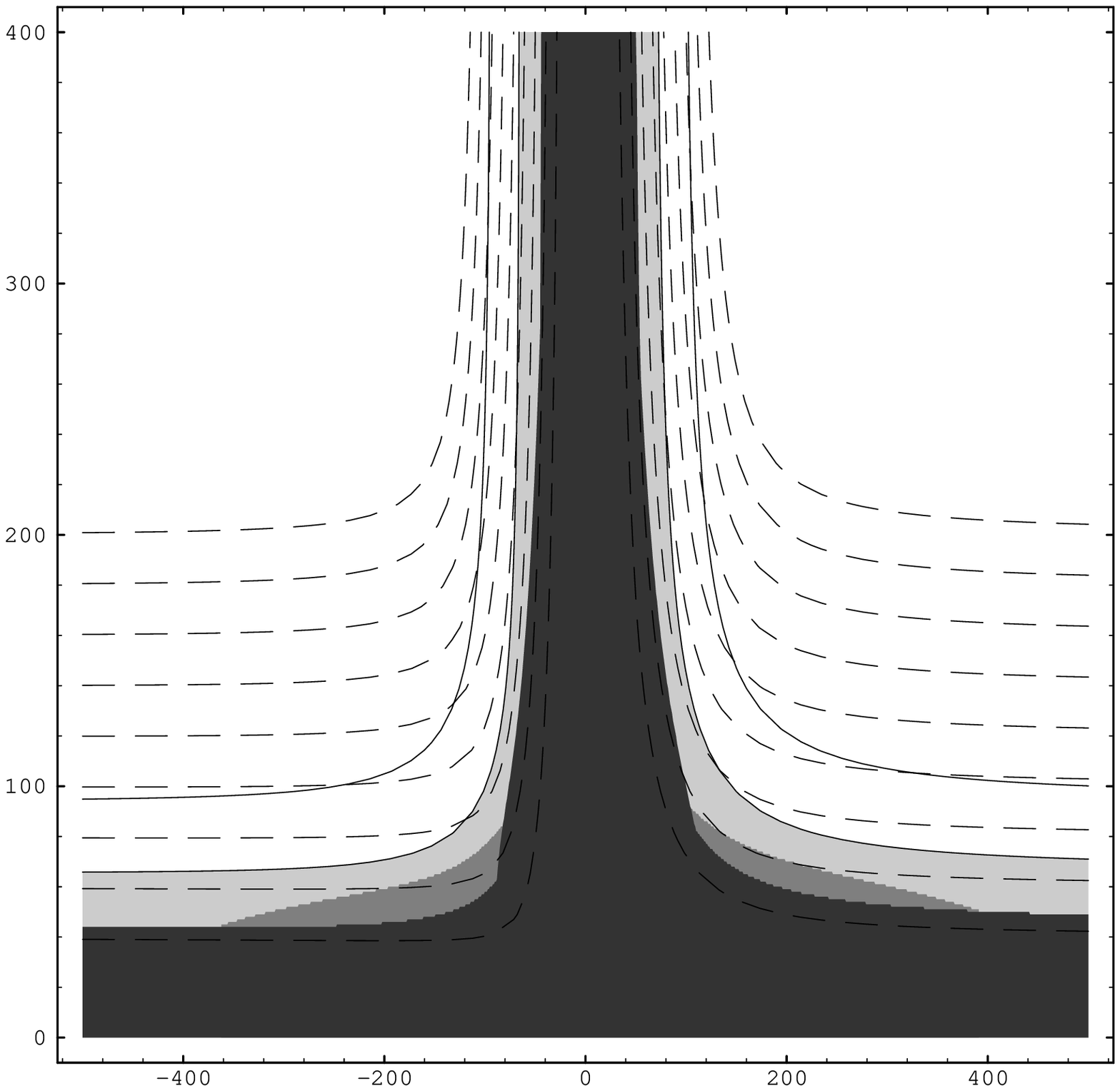}}
\put(5.2,0){$\mu/$GeV}
\put(0,4.7){\begin{sideways} $M/$GeV \end{sideways}}
\put(1.35,6){\scriptsize  $m_{\tilde{\chi}_1^0}/$GeV}
\put(1.35,1.95){\scriptsize 20}
\put(1.35,3.2){\scriptsize 50}
\put(1.35,5.45){\scriptsize 100}
\put(8.9,5.9){\scriptsize  $m_{\tilde{\chi}_1^\pm}/$GeV}
\put(9.7,2.55){\scriptsize 66.8}
\put(9.9,3.25){\scriptsize 95}
\end{picture}
\end{center}
Figure 1c:\ Contour lines for the mass of the lightest neutralino
(dashed; 20, 30, 40, 50, 60, 70, 80, 90 and 100~GeV)
and chargino (solid; 66.8 and 95~GeV) and the excluded parameter
space from LEP1 (dark: $Z$ width
measurements, gray: direct neutralino search) and LEP1.5 (bright shaded) 
in the ($M,\mu$) plane for $\tan\beta=10$.
\end{figure}
\clearpage

\setcounter{figure}{1}

\begin{figure}[p]
\centering
\setlength{\unitlength}{1cm}
\begin{picture}(14,17.8)
\put(-1.4,2.6){\includegraphics{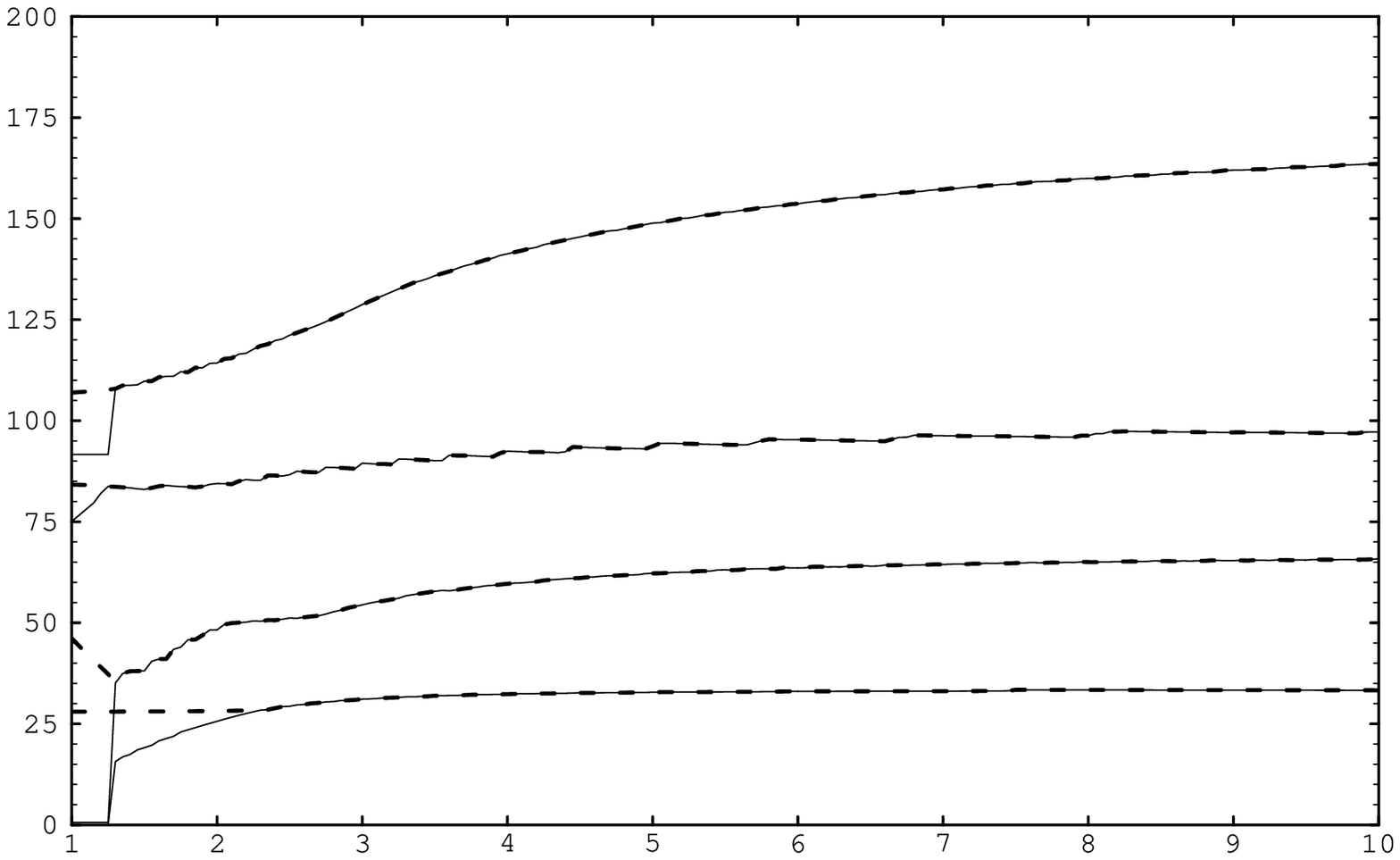}}
\put(1.8,16.9){(a)}
\put(10.0,10.4){$m_{\tilde{\chi}_1^\pm}>66.8$ GeV}
\put(7.0,9.4){$\tan\beta$}
\put(0,11.1){\begin{sideways} Neutralino mass bounds [GeV] \end{sideways}}
\put(-1.4,-6.8){\includegraphics{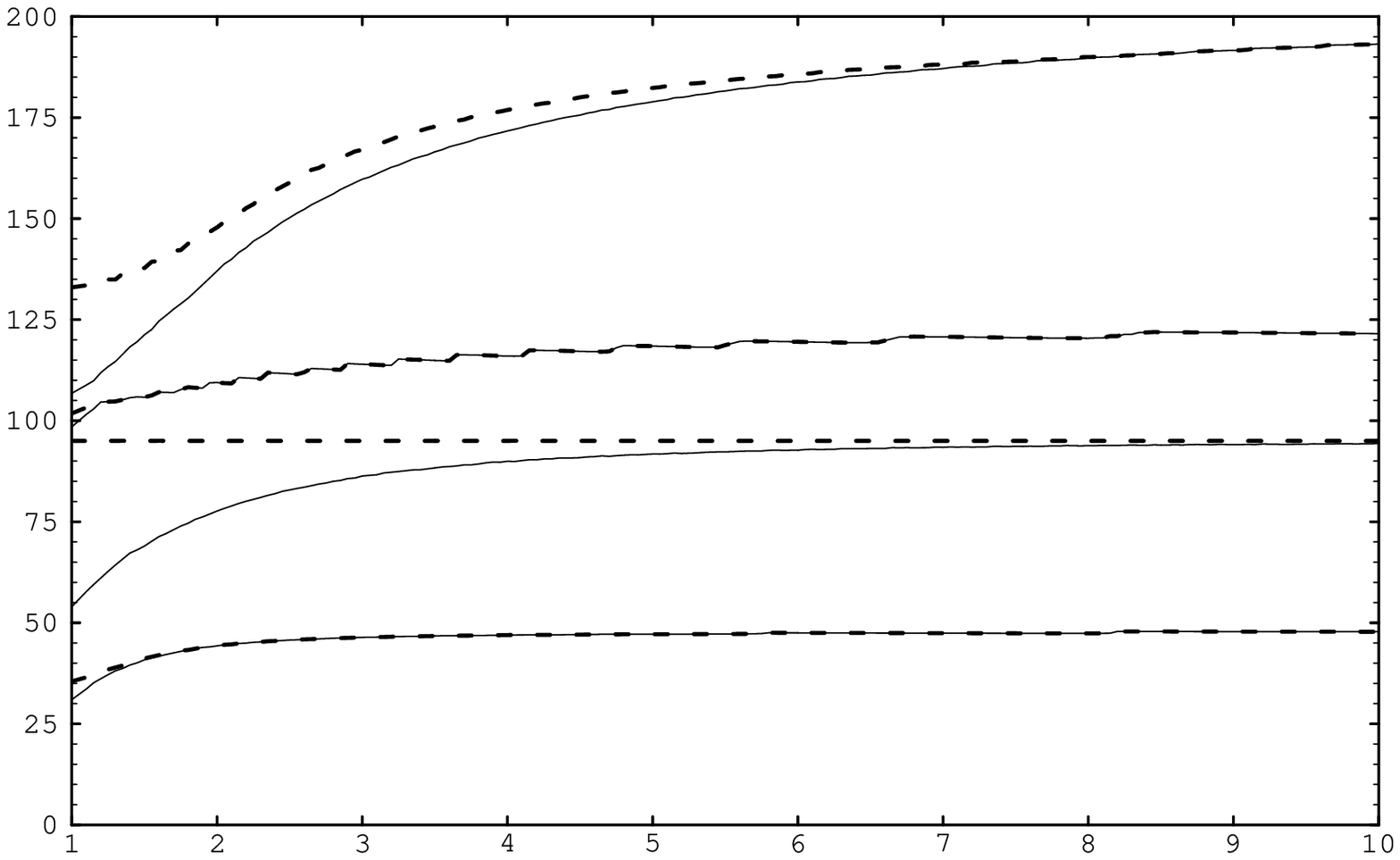}}
\put(1.8,7.5){(b)}
\put(10.0,1.0){$m_{\tilde{\chi}_1^\pm}>95$ GeV}
\put(7.0,0){$\tan\beta$}
\put(0,1.7){\begin{sideways} Neutralino mass bounds [GeV] \end{sideways}}

\end{picture}

\caption{The lower limits on the masses of the four neutralinos as a
function of $\tan\beta$ for the chargino mass bounds $m_{\tilde{\chi}_1^\pm}
>66.8$ GeV (a) and $m_{\tilde{\chi}_1^\pm}>95$ GeV (b). The dashed
lines in (a) include the CDF bound $M >50$ GeV, 
whereas in (b) they
mark the mass limits if additionally the second lightest
neutralino is found to be heavier than 95~GeV.}
\end{figure}

\clearpage
\newpage
\begin{figure}[p]

\centering
\setlength{\unitlength}{1cm}
\begin{picture}(14,17.8)
\put(-1.4,2.6){\includegraphics{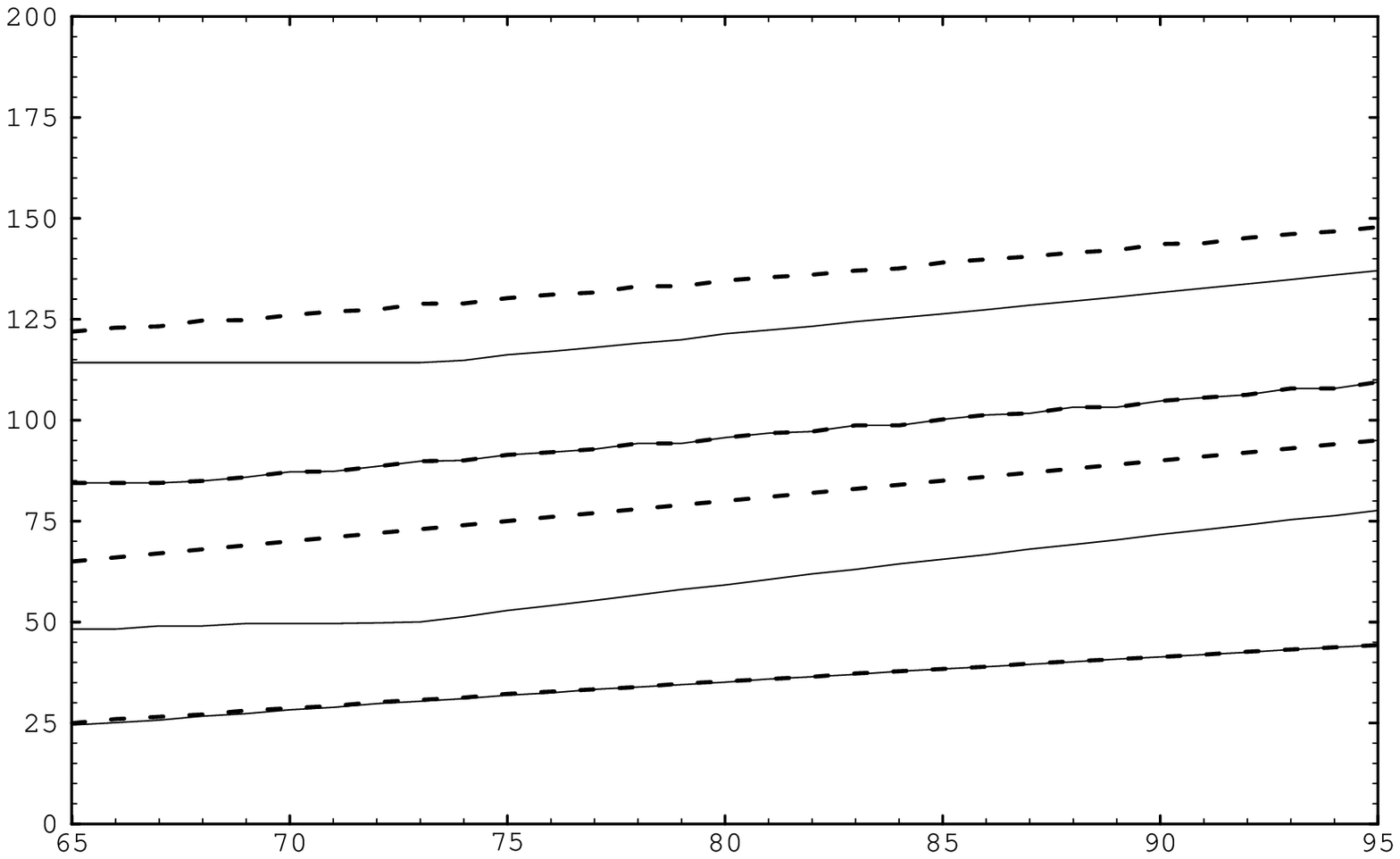}}
\put(1.8,16.9){(a)}
\put(11.3,10.5){$\tan\beta=2$}
\put(4.9,9.3){Chargino mass bound [GeV]}
\put(0,11.1){\begin{sideways} Neutralino mass bounds [GeV] \end{sideways}}
\put(-1.4,-6.8){\includegraphics{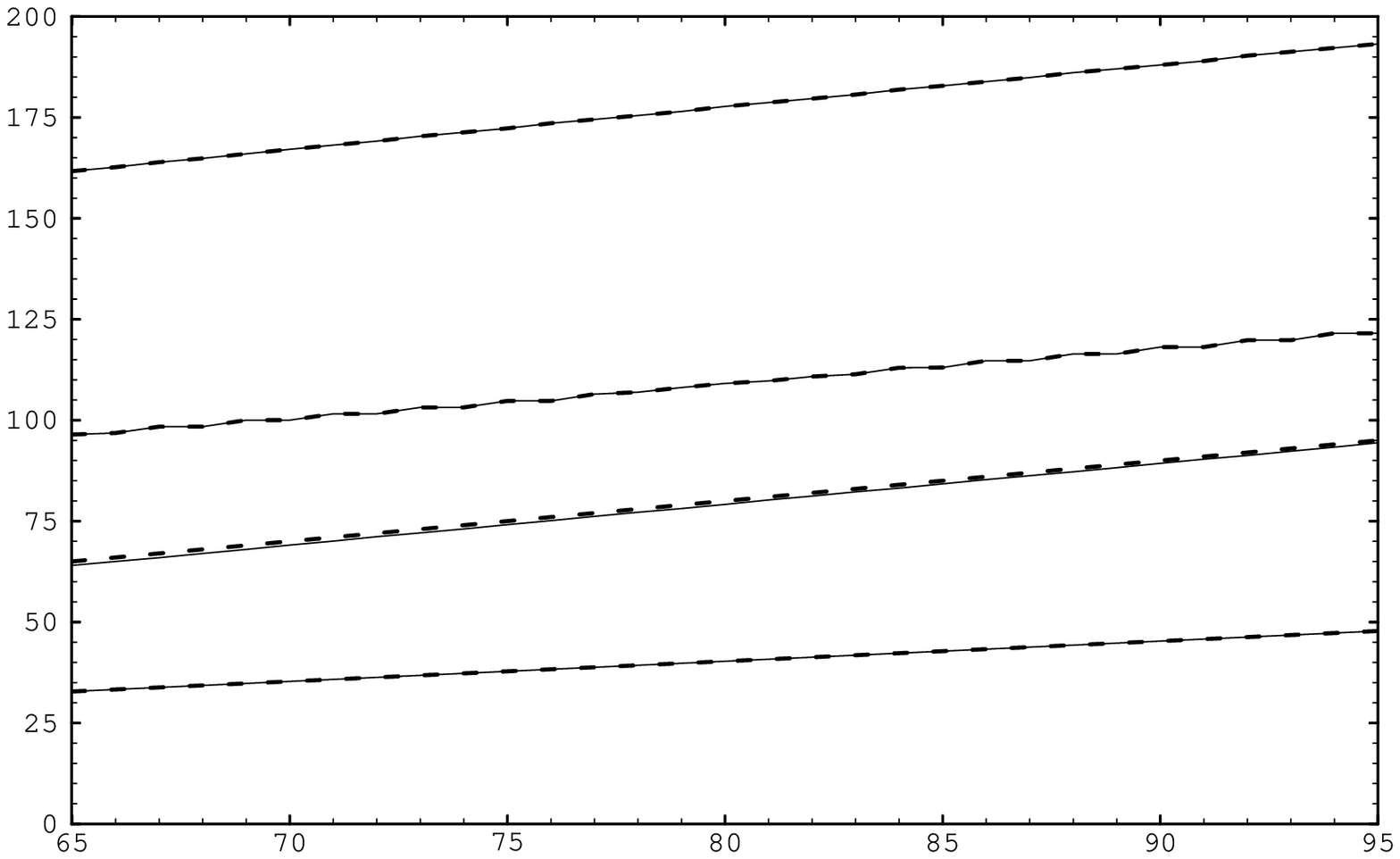}}
\put(1.8,7.5){(b)}
\put(11.3,1.1){$\tan\beta=10$}
\put(4.9,-0.1){Chargino mass bound [GeV]}
\put(0,1.7){\begin{sideways} Neutralino mass bounds [GeV] \end{sideways}}

\end{picture}

\caption{The lower limits on the masses of the four neutralinos as a
function of the lower chargino mass bound for $\tan\beta =2$ (a)
and $\tan\beta=10$ (b). The dashed lines are valid if there exists the same
mass bound for the lightest visible neutralino 
$\tilde{\chi}^0_2$.}
\end{figure}
\end{document}